\documentclass[12pt,draftcls,onecolumn]{IEEEtran}
\usepackage{graphicx}
\usepackage{amssymb}
\usepackage{amsthm}
\usepackage{hyperref}
\usepackage[cmex10]{amsmath}
\usepackage[nice]{nicefrac}
\usepackage{cite}

\DeclareMathOperator*{\argmax}{\arg\!\max}
\hypersetup
{
    colorlinks=false,
    pdfborder={0 0 0},
}
\usepackage[cmex10]{amsmath}
\begin{document}

\title{Transmit Antenna Selection in Underlay Cognitive Radio Environment}%
\author{Muhammad~Hanif, Hong-Chuan Yang,~\IEEEmembership{Senior Member,~IEEE}, and Mohamed-Slim Alouini,~\IEEEmembership{Fellow,~IEEE}%
\thanks{M. Hanif and H.-C. Yang are with the Department
of Electrical and Computer Engineering, University of Victoria (UVic), BC,
Canada (email: \{mhanif, hy\}@uvic.ca).}%
\thanks{M.-S. Alouini is with the Computer, Electrical, and Mathematical Science and Engineering (CEMSE) Division, King Abdullah University of Science and Technology (KAUST), Thuwal, Makkah Province, Saudi Arabia (email: slim.alouini@kaust.edu.sa).}%
\thanks{H.-C. Yang and M.-S. Alouini are also members of the KAUST strategic research initiative (SRI) in Uncertainty Quantification in Science and Engineering.}%
}

\maketitle
\begin{abstract}
Cognitive radio (CR) technology addresses the problem of spectrum under-utilization. In underlay CR mode, the secondary users are allowed to communicate provided that their transmission is not detrimental to primary user communication. Transmit antenna selection is one of the low-complexity methods to increase the capacity of wireless communication systems. In this article, we propose and analyze the performance benefit of a transmit antenna selection scheme for underlay secondary system that ensures the instantaneous interference caused by the secondary transmitter to the primary receiver is below a predetermined level. Closed-form expressions of the outage probability, amount of fading, and ergodic capacity for the secondary network are derived. Monte-carlo simulations are also carried out to confirm various mathematical results presented in this article.
\end{abstract}
\begin{IEEEkeywords}
Underlay cognitive radio, antenna selection, order statistics, outage probability, amount of fading, ergodic capacity.
\end{IEEEkeywords}
\section{Introduction}
\IEEEPARstart{O}{pportunistic} communication or cognitive radio (CR) technology tries to alleviate the problem of radio spectrum under-utilization. The concept of cognitive radio was first introduced in 1999 by J. Mitola \cite{MitolaCognitive}. Since then, significant amount of research has been carried on the design and performance analysis of CR systems. Three existing paradigms of CR implementation are interweave, overlay and underlay modes \cite{GoldsmithCognitive}. The interweave mode of operation depends on the presence of the space-time-frequency voids, also known as \textit{spectrum holes}. The cognitive user monitors the activity of the primary users (PUs) and transmits its information over the spectrum holes. Although the interweave mode of operation results in high spectrum efficiency but the difficulty in timely detecting the spectrum holes limits its applicability \cite{CognitiveSensingProb}. In the overlay and underlay paradigms, the secondary user (SU) transmits data even if PUs are communicating. The secondary network has to ensure that the interference caused by the secondary transmitter (ST) to the primary receiver (PR) does not exceed some fixed level determined from, for example, the quality of service (QoS) requirement of PUs \cite{GoldsmithCognitive,HaykinTutorial}. In the overlay mode of operation, the SU limits the interference caused by the ST by relaying the PU data to the PR along with its own information. The requirement of non-causal knowledge of PU data and message code-books renders it difficult to practically implement such systems \cite{GoldsmithCognitive}. In underlay mode of operation, on the other hand, the ST does not require the knowledge of the PU data for its data transmission.

Various methodologies have been proposed in the literature to control the interference caused by the ST to the PR in underlay CR systems. The most intuitive method is to adapt the power of ST in order to meet interference level requirement and maximize the CR system capacity \cite{OptPowerAllocCognitive} while satisfying its average or peak transmit power constraints. Another method is to utilize multiple antennas and perform beamformed transmission at the ST under the interference constraints \cite{BeamformingMIMOCognitive}. Multiple input multiple output (MIMO) systems can also be used in underlay CR systems to meet the interference constraint at the PR \cite{MIMOCognitive}. Although the above mentioned solutions can effectively control the interference level to the PR, they tend to result in high hardware costs due to the requirement of variable power gains for power adaptation and the use of multiple RF chains for beam-forming transmission.

Antenna selection can be used to reduce hardware complexity of multiple RF chains in a wireless transmitter \cite{MISOSelecScheme,ratioSelection,DiffAntennaSel,SEPBasedSelection}. As such, several transmit antenna selection schemes have been proposed for underlay secondary transmission. In \textit{unconstrained transmit antenna selection} (UC), the antenna leading to highest instantaneous secondary channel gain is selected to maximize the capacity of secondary network \cite{MISOSelecScheme,DrAlouiniDigCommBook}. The \textit{minimum interference selection} (MI) scheme tries to alleviate the problem of excessive interference to the PR in the unconstrained selection scheme by choosing the antenna that results in minimum interference to the PR \cite{MISOSelecScheme}. The \textit{maximum signal-to-leak interference ratio (MSLIR)} scheme tries to increase the overall system capacity by selecting the ST antenna which maximizes the ratio of ST to secondary receiver (SR) and ST to PR channel gains \cite{MISOSelecScheme,ratioSelection}. The authors in \cite{DiffAntennaSel} introduced \textit{difference antenna selection} (DS) scheme that outperforms the maximum SLIR by selecting the ST antenna leading to largest ST to SR and ST to PR channel gains weighted difference. in many scenarios while satisfying the interference constraint at PR with power adaptation. Recently, another scheme that chooses ST antenna based on minimization of symbol error probability (SEP) is proposed in \cite{SEPBasedSelection}. These schemes cannot ensure that the interference interference caused by ST to PR is below a certain threshold level unless by adopting power adaptation.

In this paper we propose a simple transmit antenna selection scheme for the secondary system that tries to maximize the instantaneous capacity of the secondary network while guaranteeing the instantaneous interference requirement of the primary network without employing power adaptation. In particular, the ST antenna that leads to the best ST to SR channel quality while still satisfying the interference requirement at the PR is used for transmission with constant power. We develop exact statistical characterization of the  instantaneous signal-to-interference plus noise (SINR) at the SR with the consideration of the interference from PT transmission. Closed-form expressions of the outage probability, amount of fading and ergodic capacity are also presented. Monte-carlo simulations are carried out to validate the derived analytical results. Through the selected numerical examples we show that the proposed scheme can outperform other antenna selection schemes while satisfying the hard interference limit at the PR.

Rest of the paper is organised as follows. System model along with the proposed scheme is introduced in Section \ref{sec:SysModel}.  The mathematical expressions for the performance metrics for the proposed scheme are presented in Section \ref{sec:PerfAnalysis} which is followed by the conclusion of the paper.

\section{System Model}
\label{sec:SysModel}
\begin{figure}[ht]
  \centering
  \includegraphics{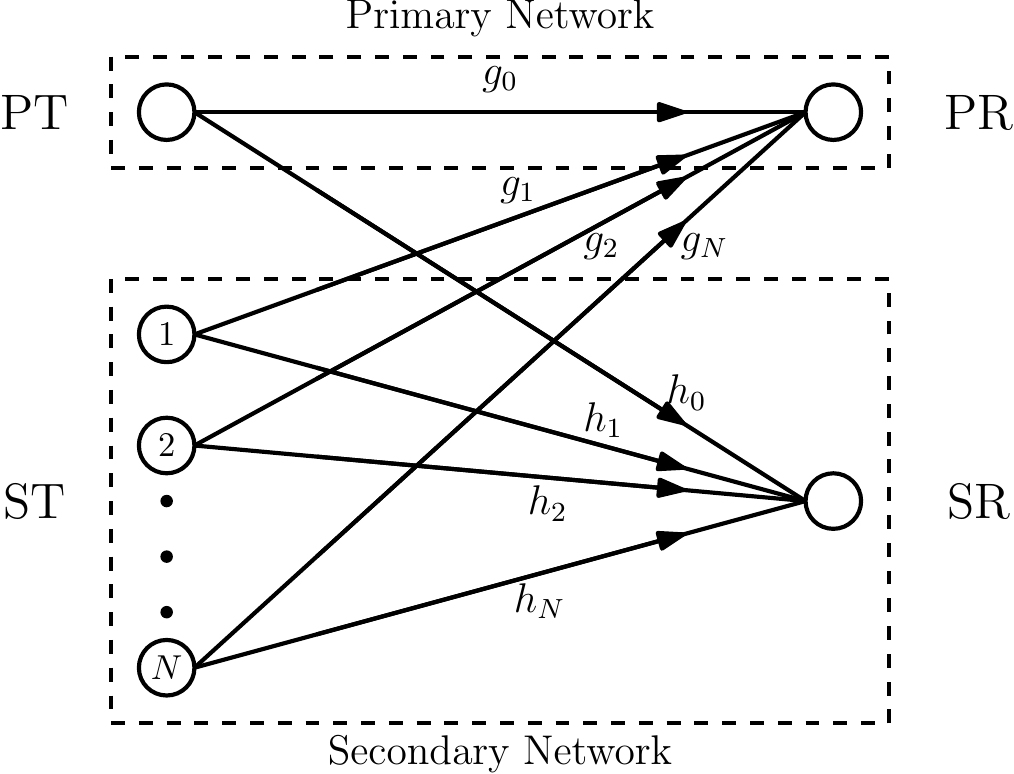}\\
  \caption{Underlay cognitive radio setup with $N$ ST antennas.}\label{CRTransSel}
\end{figure}

Consider an underlay CR setup as shown in Fig. \ref{CRTransSel}. Here, for the sake of simplicity, the primary transmitter (PT), PR, and SR are all equipped with only single antenna while the ST is equipped with $N$ antennas. We use $h_0$ to denote the complex channel gains from PT to SR, $g_i$ that from the $i$th ST antenna to PR, and $h_i$ that from the $i$th ST antenna to SR where $i=1, 2, \cdots, N$. Under independent and identically distributed (IID) Rayleigh fading channel model, the probability density functions (PDFs) of $|h_0|^2$, $|h_i|^2$, and $|g_i|^2$ are $f_{h_0}(x)=\lambda_{ps} e^{-\lambda_{ps} x}u(x)$, $f_h(x)=\lambda_{ss} e^{-\lambda_{ss} x}u(x)$, and $f_g(x)=\lambda_{sp} e^{-\lambda_{sp} x}u(x)$  respectively where the rate parameters $\lambda_{ps}, \lambda_{ss}$, and $\lambda_{sp}$ are all positive reals and $u(x)$ is the unit step function. Also, the cumulative distribution functions (CDFs) of $|h_0|^2$, $|h_i|^2$, and $|g_i|^2$ are denoted by $F_{h_0}(x)$, $F_h(x)$, and $F_g(x)$ respectively. The transmit power at PT and ST are $P_M$ and $P_S$ respectively while the theshold $T$ denotes the interference power limit at the PR, which can be directly related to the interference temperature constraint. Lastly, the power of zero mean white Gaussian noise at the SR will be denoted by $N_0$. With the defined notation, the difference antenna selection scheme, for example, can be described as the selection of $i^*$th ST antenna such that
\begin{IEEEeqnarray}{c}
i^*=\argmax_{i=1,2\cdots,N}\{\eta|h_i|^2-(1-\eta)|g_i|^2\},
\end{IEEEeqnarray}
where the selection weight, $\eta \in [0,1]$, can be found by maximizing the mutual information or minimizing the outage probability of the secondary network \cite{DiffAntennaSel}. Note that the cases $\eta=0$ and $\eta=1$ correspond to the minimum interference and unconstrained selection schemes described above respectively.

We propose an antenna selection scheme that ensures the instantaneous interference caused by the ST to the PR is below an acceptable level while maximizing the instantaneous SINR at the SR. In particular, we choose the  `best' ST antenna, in terms of achieving the highest SINR at SR, that satisfies the interference power constraint. If none of the antennas satisfy the interference power constraint at PR, then we hold the transmission for a channel coherence time and check the channel condition again. This scheme can be implemented in the following steps.
\begin{enumerate}
  \item Sort out the square of absolute channel gains from ST antennas to SR, $|h_i|^2$, in decreasing order. Let the sorted squared absolute channel gains be denoted as $|h_{(N)}|^2 \ge |h_{(N-1)}|^2 \ge \cdots \ge |h_{(1)}|^2$.
  \item Starting from the antenna with gain $|h_{(N)}|^2$, choose the antenna that satisfies the interference power constraint at the PR for data transmission.
  \item If none of the ST antennas satisfy the constraint at PR, then halt the transmission for a channel coherence time and repeat steps 1 and 2.
  \end{enumerate}
This scheme will generally result in selection of the $k$th best ST to SR channel, $h_{(N-k+1)}$, for $k=1, 2, \cdots, N$.
\section{Performance Analysis}
\label{sec:PerfAnalysis}
We now derive the statistics of the received SINR with the proposed antenna selection scheme.
\subsection{Statistics of Received SINR at SR}
Let $\xi_k$ denote the received SINR at the SR when $k$th best ST antenna is selected. Mathematically, we have
\begin{IEEEeqnarray}{c}
\xi_k = \frac{P_S|h_{(N-k+1)}|^2}{N_0+P_M|h_0|^2}.
\end{IEEEeqnarray}
The PDF of SINR, $\xi_k$, can be computed by conditioning on $|h_0|^2$ as
\begin{IEEEeqnarray}{rl}
\label{genXiPDF}
f_{\xi_k}(x)&=\!\int_0^\infty{f_{h_{(N-k+1)}}\left(\frac{xP_My}{P_S}+\frac{N_0 x}{P_S}\right)\cdot \left(\frac{P_My}{P_S}+\frac{N_0}{P_S}\right) \cdot f_{h_0}(y)dy},\IEEEeqnarraynumspace
\end{IEEEeqnarray}
where $f_{h_{(N-k+1)}}(x)$ is the PDF of $|h_{(N-k+1)}|^2$ given as \cite{DavidBook}
\begin{IEEEeqnarray}{c}
\label{kthBestPDF}
f_{h_{(N-k+1)}}(x) = k\binom{N}{k} \lambda_{ss} e^{-\lambda_{ss} k x} \left(1-e^{-\lambda_{ss} x}\right)^{N-k}u(x),\IEEEeqnarraynumspace
\end{IEEEeqnarray}
Using the \eqref{integralOne} and \eqref{integralTwo} given in Appendix \ref{sec:IntRelations}, one can easily derive the PDF of $\xi_k$ as
\begin{IEEEeqnarray}{rl}
\label{XikPDF}
f_{\xi_k}(x)=&\frac{k\lambda_{ss} \lambda_{ps} e^{-\left(\frac{k\lambda_{ss} N_0}{P_S}\right)x}}{(k\lambda_{ss} P_M x + \lambda_{ps} P_S)^2}\binom{N}{k}\Bigg[
\resizebox{0.6\hsize}{!}{$P_M P_S \cdot{}_3F_2\left(\begin{array}{c}
                 k+\frac{\lambda_{ps} P_S}{\lambda_{ss} P_M x}, k+\frac{\lambda_{ps} P_S}{\lambda_{ss} P_M x}, k-N \\
                 k+1+\frac{\lambda_{ps} P_S}{\lambda_{ss} P_M x}, k+1+\frac{\lambda_{ps} P_S}{\lambda_{ss} P_M x} \\
               \end{array}
;e^{-\frac{\lambda_{ss} N_0 x}{P_S}}\right)$} \IEEEnonumber \\
& +{}N_0 (kx\lambda_{ss} P_M + \lambda_{ps} P_S)\cdot
{}_2F_1\left(\begin{array}{c}
                 k+\frac{\lambda_{ps} P_S}{\lambda_{ss} P_M x}, k-N \\
                 k+1+\frac{\lambda_{ps} P_S}{\lambda_{ss} P_M x} \\
               \end{array}
;e^{-\frac{\lambda_{ss} N_0 x}{P_S}}\right)\Bigg],
\end{IEEEeqnarray}
where ${}_2F_1\left(\begin{matrix}m,n\\p\end{matrix};q\right)$ and ${}_3F_2\left(\begin{matrix}m,n,o\\p,q\end{matrix};r\right)$ are the Hypergeometric function generalized Hypergeometric function respectively \cite{IntegralTables}. The PDF of $\xi_k$ can also be derived by expanding \eqref{kthBestPDF} using binomial theorem. Performing integration as in \eqref{genXiPDF}, we obtain
\begin{IEEEeqnarray} {rl}
\label{XikPDFAlternative}
f_{\xi_k}(x) = k\binom{N}{k}\lambda_{ss}\lambda_{ps} \sum_{j=0}^{N-k}{(-1)^j \binom{N-k}{j}} e^{-\frac{N_0 x \lambda_{ss} (k+j)}{P_S}} \Bigg( & \frac{N_0}{P_S\lambda_{ps}+\lambda_{ss}(k+j)P_M x} \IEEEnonumber \\
& +\frac{P_S P_M }{(P_S\lambda_{ps}+\lambda_{ss}(k+j)P_M x)^2}\Bigg).\IEEEeqnarraynumspace
\end{IEEEeqnarray}
\subsection{Outage Probability}
Outage at the SR occurs whenever the SINR at the SR falls below a threshold $\xi_T$. Let $p_k$ be the probability of selecting the $k$th best ST antenna for data transmission. Also, let $p_\phi$ be the probability of the event that none of the channel gains satisfy the constraint $P_S |g_i|^2\le T$. Then, it can easily be shown that
\begin{IEEEeqnarray}{c}
\label{selProb}
p_k = (1-F_g(T/P_S))^{k-1}F_g(T/P_S)=e^{-(k-1)\lambda_{sp} T/P_S}(1-e^{-\lambda_{sp} T/P_S}),~~~k=1, 2, \cdots, N,
\end{IEEEeqnarray}
and
\begin{IEEEeqnarray}{c}
\label{outProb}
p_{\phi}=(1-F_g(T/P_S))^N=e^{-N\lambda_{sp} T/P_S}.
\end{IEEEeqnarray}
So, the probability of outage can be computed as
\begin{IEEEeqnarray}{c}
P_{o}=p_{\phi}+\sum_{k=1}^{N}{p_k F_{\xi_k}(\xi_T)},
\end{IEEEeqnarray}
where $p_k$ and $p_\phi$ are defined in \eqref{selProb} and \eqref{outProb} respectively and $F_{\xi_k}(x)$ is the CDF of $\xi_k$ which can be computed by using the following relationship \cite{DavidBook}
\begin{IEEEeqnarray}{c}
F_{h_{(N-k+1)}}(x) = \sum_{i=0}^{k-1}{\binom{N}{i}\left(1-e^{-\lambda_{ss} x}\right)^{N-i}e^{-\lambda_{ss} ix}}u(x),
\end{IEEEeqnarray}
 and \eqref{integralOne} in Appendix \ref{sec:IntRelations} to obtain
\begin{IEEEeqnarray}{l}
\label{XikCDF}
F_{\xi_k}(x)=\lambda_{ps} P_S \sum_{i=0}^{k-1}{\begin{pmatrix}N\\ i\end{pmatrix} \exp{\left(\frac{-N_0 i\lambda_{ss} x}{P_S}\right)}}\cdot
{\frac{{}_2F_1\left(\begin{array}{c}
                 i+\frac{\lambda_{ps} P_S}{\lambda_{ss} P_M x}, i-N \\
                 i+1+\frac{\lambda_{ps} P_S}{\lambda_{ss} P_M x} \\
               \end{array}
;e^{-\frac{\lambda_{ss} N_0 x}{P_S}}\right)}{\lambda_{ps} P_S+\lambda_{ss} P_M i x}}.
\end{IEEEeqnarray}

\begin{figure}[ht]
  \centering
  \includegraphics[width=\columnwidth]{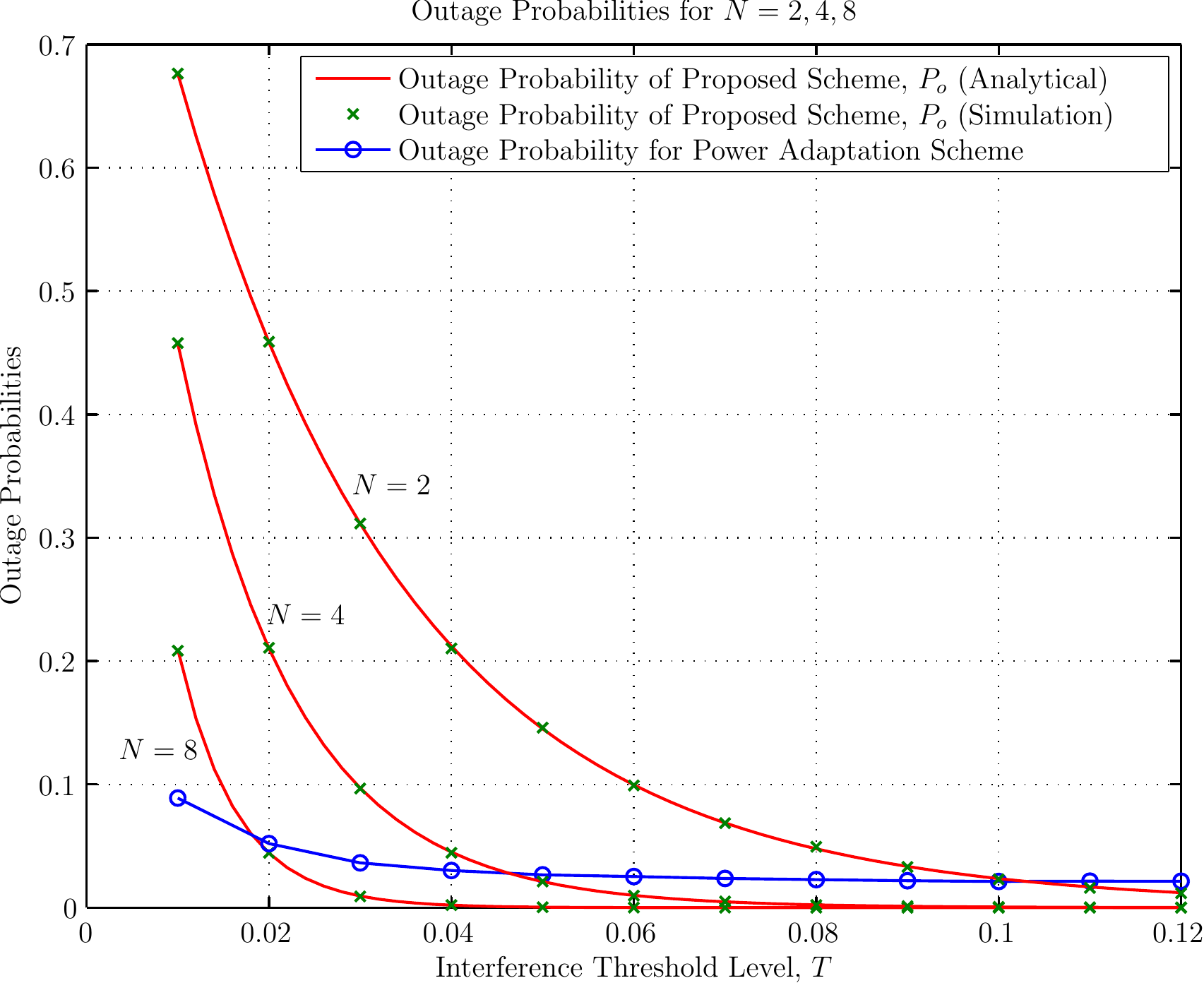}\\
  \caption{Outage probability of single-antenna and proposed antenna selection schemes against interference threshold level, $T$.}\label{OutageProbCurve}
\end{figure}

Fig. \ref{OutageProbCurve} shows the trend of outage probability as the interference threshold level, $T$, is increased. The simulation parameters are $P_M=1$, $P_S=0.5$, and $N_0=10^{-2}$. Also, the fading parameters are $\lambda_{ps}=\lambda_{sp}=10$ and $\lambda_{ss}=3$ while $N=2, 4, 8$. Here Monte-carlo simulation results for $10^5$ trials are also presented. Note that, as expected, the chances of outage is reduced by increasing the number of antennas at the ST. Also note that when the threshold is small enough (or equivalently when $P_S$ is high), then the outage probability increases rapidly. Fig. \ref{OutageProbCurve} also shows the outage probability of single-antenna with power adaptation scheme, in which the ST adapts its transmit power, $p$, depending on ST to PR as $p=\min{\left(P_S,T/|g_1|^2\right)}$. Therefore, the outage probability of the single-antenna power adaptation scheme for small $T$ is smaller than that of the proposed scheme. However, increasing the number of transmit antennas, $N$, for the proposed scheme improves the outage performance dramatically.
\subsection{Higher Order Amount of Fading}
The higher order amount of fading, $\mathrm{AF}^{(n)}$, is defined as
\begin{IEEEeqnarray}{c}
\mathrm{AF}^{(n)}=\frac{E\left[\xi^n\right]}{E\left[\xi\right]^n}-1,
\end{IEEEeqnarray}
where $\xi$ is the end-to-end SINR\footnote{Note that in the definition of $\mathrm{AF}^{(n)}$ SINR, $\xi$, is used rather than the SNR as defined in \cite{AsymptoticLowerBound}.}. For the proposed scheme, $\mathrm{AF}^{(n)}$ specializes to
\begin{IEEEeqnarray}{c}
\mathrm{AF}^{(n)}=\frac{\sum_{k=1}^{N}{p_k E[\xi_k^n]}}{\left(\sum_{k=1}^{N}{p_k E[\xi_k]}\right)^n}-1,
\end{IEEEeqnarray}
where $p_k$ is defined in \eqref{selProb}. Using relationships in Appendix \ref{sec:SINRMoment}, the $n$th moment of $\xi_k$ can be computed as
\begin{IEEEeqnarray}{c}
\label{XiktonExpect}
E\left[\xi_k^n\right] = n! \left(\frac{\lambda_{ps} P_S}{\lambda_{ss} P_M}\right)^n e^{\lambda_{ps} N_0/P_M} \Gamma\left(1-n,\frac{\lambda_{ps} N_0} {P_M}\right) \sum_{l=k}^{N}{\frac{1}{l^{n+N-k}\prod_{\substack{j=k\\j\neq l}}^{N}{\left(\frac{1}{l}-\frac{1}{j}\right)}}},
\end{IEEEeqnarray}
and
\begin{IEEEeqnarray}{c}
\label{XikExpect}
E[\xi_k]=\frac{\lambda_{ps} P_S}{\lambda_{ss} P_M}e^{\frac{\lambda_{ps} N_0}{P_M}}E_1\left(\frac{\lambda_{ps} N_0}{P_M}\right)\left(H_N-H_{k-1}\right),
\end{IEEEeqnarray}
where $\Gamma(a,z)=\int_z^{\infty}{t^{a-1} e^{-t} dt}$ is the incomplete Gamma function, $E_1(z)=\int_{z}^{\infty}{\frac{e^{-t}}{t} dt}=\Gamma(0,z)$ is the exponential integral function, and $H_m$ is the Harmonic number \cite{handbookSpecialFunc}. Using \eqref{XiktonExpect} and \eqref{XikExpect}, the $\mathrm{AF}^{(n)}$ can be computed easily. Note that the $\mathrm{AF}^{(n)}$ involves only standard functions which can easily be computed in standard softwares like MATLAB and Mathematica.

\begin{figure}[ht]
  \centering
  \includegraphics[width=\columnwidth]{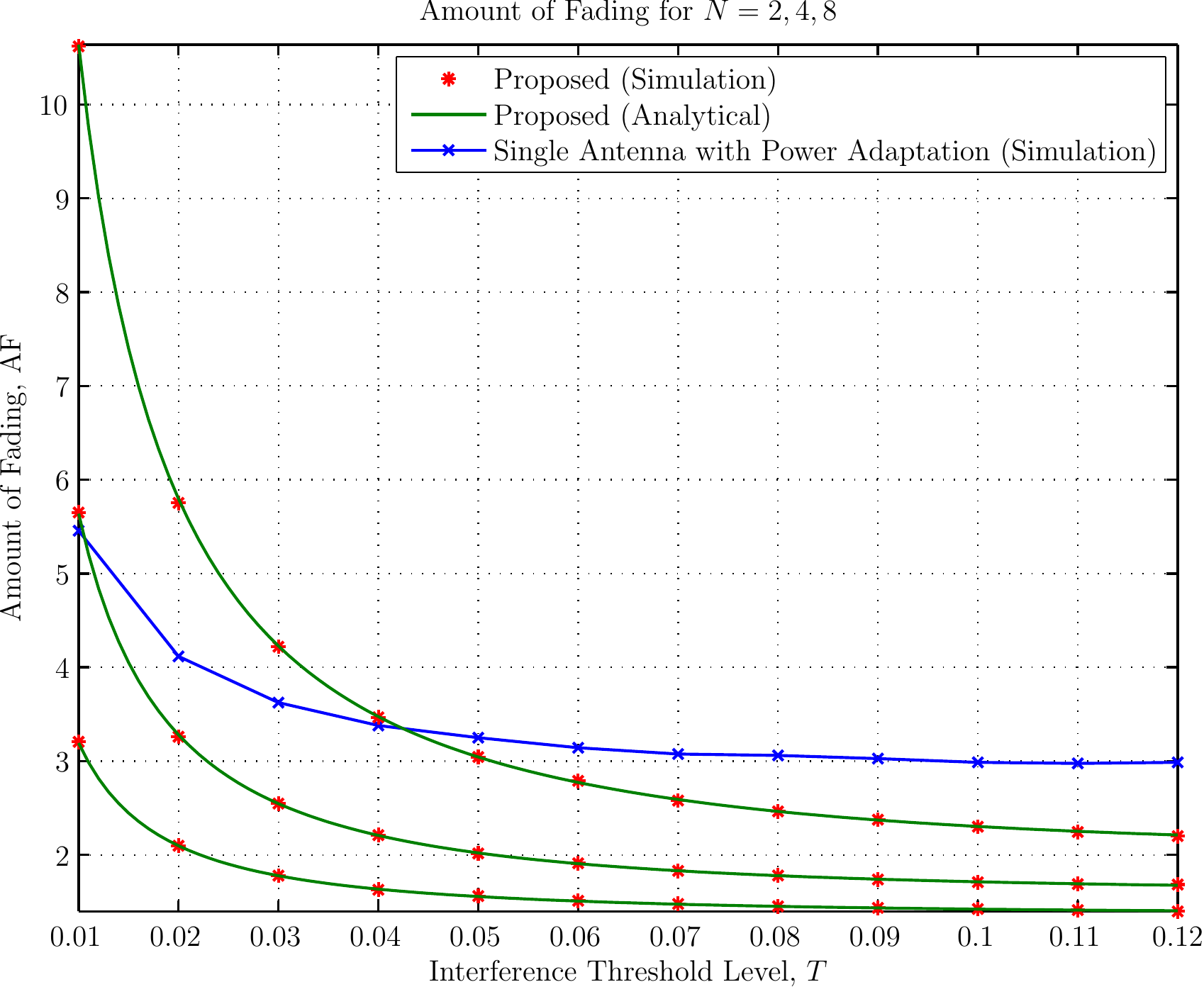}\\
  \caption{Amount of fading of single-antenna and proposed antenna selection schemes against interference threshold level, $T$.}\label{AFComparison}
\end{figure}

Fig. \ref{AFComparison} shows the trend of amount of fading, $\mathrm{AF}=\mathrm{AF}^{(2)}$, as the interference threshold level, $T$, is varied. The simulation parameters are $P_M=1$, $P_S=0.5$, $N_0=10^{-2}$. The fading parameters are $\lambda_{ps}=\lambda_{sp}=10$ and $\lambda_{ss}=3$ while $N=2, 4, 8$. Note that, as expected, the AF is reduced by increasing the threshold, $T$. This is because for high values of $T$, the best antenna is selected for most of the time and hence the variation in the SINR is reduced. While for the small values of $T$, probability of selection of poor channels (whose variance is higher than the normal channel's even) is also increased. Hence the AF becomes larger for small values of $T$.
\subsection{Ergodic Capacity}
Ergodic capacity for the proposed scheme can be computed as $\overline{C}=\sum_{k=1}^N{p_k \overline{C}_k}$ where $\overline{C}_k$ is the ergodic capacity for the $k$th best path and $p_k$ is defined in \eqref{selProb}. Using \eqref{XikPDFAlternative}, and \eqref{exactCapHelper} in Appendix \ref{sec:IntRelErgodicCap} we derive the closed-form expression of $\overline{C}_k$ as
\begin{IEEEeqnarray}{c}
\label{exactCkCap}
\overline{C}_k = \frac{k\binom{N}{k}\lambda_{ps}}{\ln{2}} \sum_{j=0}^{N-k}{(-1)^j \binom{N-k}{j}} \frac{P_S\Big[e^{\frac{N_0 \lambda_{ps}}{P_M}}E_1\Big(\frac{N_0 \lambda_{ps}}{P_M}\Big) - e^{\frac{N_0 \lambda_{ss}(k+j)}{P_S}}E_1\Big( \frac{N_0 \lambda_{ss}(k+j)}{P_S} \Big)\Big]}{(k+j)[P_M \lambda_{ss}(k+j) - P_S \lambda_{ps}]}.\IEEEeqnarraynumspace
\end{IEEEeqnarray}
Here $E_1(z)=\int_{z}^{\infty}{\frac{e^{-t}}{t} dt}$ is the exponential integral function.

\begin{figure}[ht]
  \centering
  \includegraphics[width=\columnwidth]{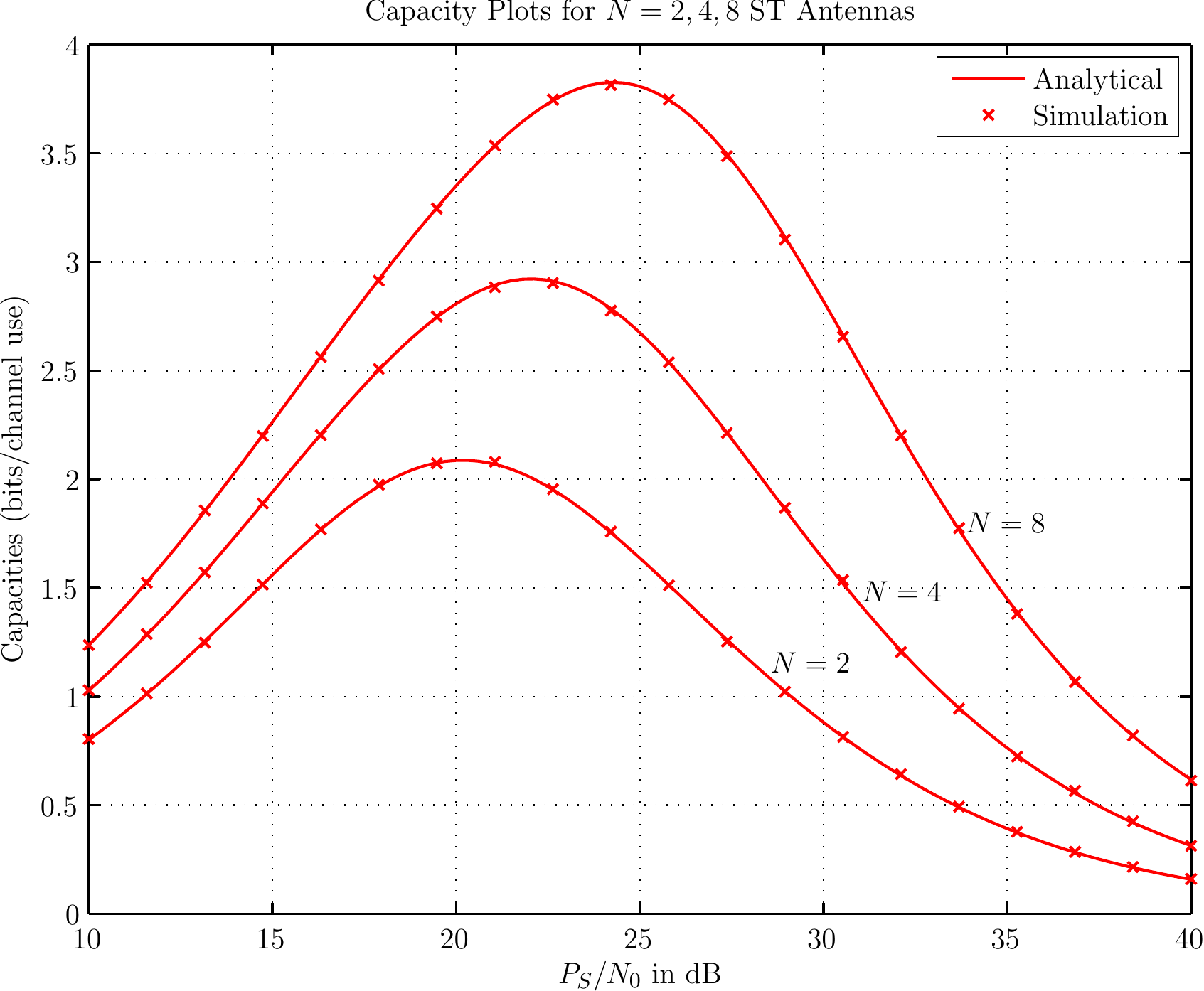}\\
  \caption{Capacity of proposed antenna selection scheme for $N=2, 4, 8$ ST antennas.} \label{N248Case}
\end{figure}

Fig. \ref{N248Case} shows the capacity of the proposed scheme for different values of $N$. The system parameters are $\lambda_{ps}=\lambda_{sp}=10$, $\lambda_{ss}=3$, $T=0.1$, $P_M=1$, $N_0=10^{-2}$ and $P_S$ is varied from $0.1$ to $100$. Observe that the capacity increases with increasing the number of ST antennas as expected by intuition. Ergodic capacity computed using Monte-carlo simulations with $10^5$ trials also confirm our analytic results as evident in Fig. \ref{N248Case}.

\begin{figure}[ht]
  \centering
  \includegraphics[width=\columnwidth]{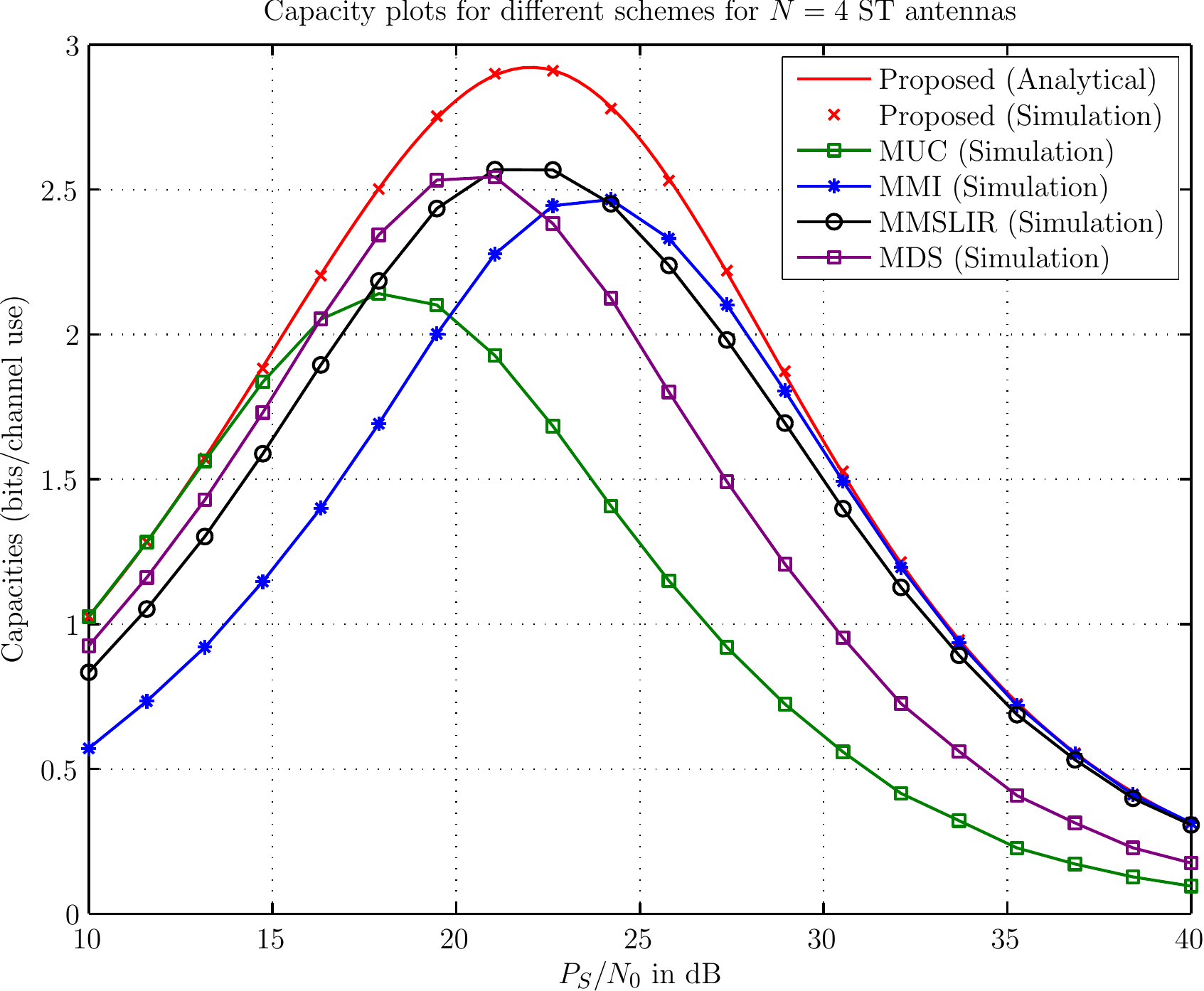}\\
  \caption{Capacity of proposed antenna selection schemes with MMI, MUC, MMSLIR, and MDS based selection rules for $N=4$ ST antennas.} \label{N4Case}
\end{figure}

In Fig. \ref{N4Case} we compare the proposed scheme with several existing antenna selection scheme by plotting ergodic capacity against normalized ST transmit power. The parameters used in simulation are $\lambda_{ps}=\lambda_{sp}=10$, $\lambda_{ss}=3$, $T=0.1$, $P_M=1$, $N=4$, $N_0=10^{-2}$ and $P_S$ is varied from $0.1$ to $100$. For fair comparison, we modified the MI, UC, MSLIR, and DS schemes such that ST stops transmission when the interference caused by transmission through selected antenna is above the threshold level, resulting in modified minimum interference (MMI), modified unconstrained (MUC), modified maximum signal-to-leak interference ratio (MMSLIR), and modified difference antenna selection (MDS) schemes \footnote{For DS scheme, the antenna that results in largest $Z_i=\eta |h_i|^2-(1-\eta)|g_i|^2$ is chosen. For the simulation purpose $\eta=0.2$.}. Observe that the proposed scheme outperforms all other schemes for all values of $P_S/N_0$. Also observe that the capacity of all schemes decrease if we increase the transmit power too much. The reason for that is the frequent occurrence of the outage event.

It is also interesting to observe that the proposed scheme incurs a relatively small feedback burden as it requires only the information about feasibility of data transmission through ST antennas. Although the feedback burden on MMI and MUC is less than that of the proposed scheme but their performance is not as good as that of the proposed scheme for complete $P_S/N_0$ range. The MMSLIR and MDS schemes, on the other hand, not only require knowledge of $|g_i|^2$ at the ST but also has worse performance as compared with the proposed scheme.

Lastly, for small values of $P_S/N_0$, the proposed scheme and the MUC scheme has similar performance because for small $P_S$, the interference caused by the ST to PR is small for all antennas and hence our scheme and MUC choose the antenna that ensures maximum data rate in the secondary network. For higher values of $P_S$, interference caused by the ST to PR becomes larger. Hence our scheme has similar performance as that of MMI.

\section{Conclusion}
Antenna selection results is a low cost method to reap the diversity benefits. In cognitive radio environment, the maximum diversity benefit might not be gained due to excessive interference to the primary receiver. The $k$th best selection scheme proposed in this paper tries to maximize the end-to-end secondary link SINR while ensuring the interference caused by the secondary transmitter to the primary receiver is below a certain level. Closed-form expressions of outage probability, amount of fading, and capacity were presented in this article.
\appendices
\section{Integration Relationships Used in Derivation of Statistics of SINR}
\label{sec:IntRelations}
First of all observe that
\begin{IEEEeqnarray}{c}
\label{integralOne}
\int_0^{\infty}{e^{-ax}(1-c e^{-bx})^d dx} =\int_0^{1}{u^{a/b-1}(1-c u)^d du} =\frac{{}_2F_1\left(\begin{matrix}a/b,~-d\\a/b+1\end{matrix};c\right)}{a},
\end{IEEEeqnarray}
Also, it is easy to show that
\begin{IEEEeqnarray}{c}
\frac{d}{d u} \left[u^m {}_2F_1\left(\begin{matrix}m,n\\m+1\end{matrix};cu\right)\right]=m u^{m-1}(1-cu)^{-n}.
\end{IEEEeqnarray}
Therefore,
\begin{IEEEeqnarray}{rl}
\int_0^{\infty}{x e^{-ax}(1-c e^{-bx})^d dx}
= &\frac{1}{ab}\int_1^{0}{\ln u~\frac{d}{du}\left( u^{a/b} {}_2F_1\left(\begin{matrix}a/b,-d\\a/b+1\end{matrix};cu\right)\right) du} \IEEEnonumber\\
=& \frac{1}{ab}\int_0^{1}{u^{a/b-1} {}_2F_1\left(\begin{matrix}a/b,-d\\a/b+1\end{matrix};cu\right)}.
\end{IEEEeqnarray}
Using the following Euler's integral relationship \cite{IntegralTables}
\begin{IEEEeqnarray}{l}
\int_{0}^{1}{u^{\nu-1}(1-u)^{\mu-1}{}_2F_1\left(\begin{matrix}m,n\\p\end{matrix};cu\right) du}
= \frac{\Gamma(\mu)\Gamma(\nu)}{\Gamma(\mu+\nu)} {}_3F_2\left(\begin{matrix}\nu,m,n\\ \mu+\nu,p\end{matrix};c\right),
\end{IEEEeqnarray}
one can easily obtain
\begin{IEEEeqnarray}{l}
\label{integralTwo}
\int_0^{\infty}{x e^{-ax}(1-c e^{-bx})^d dx} =\frac{{}_3F_2\left(\begin{matrix}\nicefrac{a}{b},\nicefrac{a}{b},~-d\\\nicefrac{a}{b}+1,\nicefrac{a}{b}+1
\end{matrix};c\right)}{a^2}.\IEEEeqnarraynumspace
\end{IEEEeqnarray}
\section{Derivation of Moments of SINR}
\label{sec:SINRMoment}
First of all observe that
\begin{IEEEeqnarray}{c}
\label{eqIndProd}
E\left[\xi_k^n\right] = P_S^n E\left[|h_{(N-k+1)}|^{2n}\right]E\left[\frac{1}{\left(N_0+P_M |h_0|^2\right)^n}\right].\IEEEeqnarraynumspace
\end{IEEEeqnarray}
It can be shown for Rayleigh fading environment
\begin{IEEEeqnarray}{rl}
E\left[\frac{1}{\left(N_0+P_M |h_0|^2\right)^n}\right] = & \lambda_{ps} \int_0^{\infty}{\frac{e^{-\lambda_{ps} x}}{\left(N_0 + P_M x\right)^n} dx}\IEEEnonumber\\
\label{denExpect}
=& \left(\frac{\lambda_{ps}}{P_M}\right)^n e^{\lambda_{ps} N_0/P_M}\Gamma\left(1-n,\frac{\lambda_{ps} N_0}{P_M}\right).
\end{IEEEeqnarray}
Note that $|h_{(N-k+1)}|^{2}$ can be written as sum of independent exponential RVs \cite{SukhatmeExpSum}. Therefore,
\begin{IEEEeqnarray}{c}
E\left[|h_{(N-k+1)}|^{2n}\right] = E\left[\left(y_N + y_{N-1} + \cdots + y_k\right)^n\right],
\end{IEEEeqnarray}
where $y_j$ is an exponential RV with parameter $j\lambda_{ss}$. Using multinomial theorem, one can write
\begin{IEEEeqnarray}{c}
E\left[|h_{(N-k+1)}|^{2n}\right] = \sum_{\sum_i{r_i}=n}\frac{n!}{r_0! r_1! \cdots r_{N-k}!} E\left[y_N^{r_0}y_{N-1}^{r_1}\cdots y_k^{r_{N-k}}\right].
\end{IEEEeqnarray}
Using the independence of $y_j$s for $j=k, k+1, \cdots, N$ and the fact that $E\left[y_j^{r_i}\right]=r_i!/(j \lambda_{ss})^{r_i}$, one gets
\begin{IEEEeqnarray}{c}
\label{numExpect}
E\left[|h_{(N-k+1)}|^{2n}\right] =\frac{n!}{\lambda_{ss}^n} \sum_{\sum_i{r_i}=n}{\prod_{j=k}^{N}j^{-r_{N-j}}}
=\frac{n!}{\lambda_{ss}^n} \sum_{l=k}^{N}{\frac{1}{l^{n+N-k}\prod_{\substack{j=k\\j\neq l}}^{N}{\left(\frac{1}{l}-\frac{1}{j}\right)}}}.
\end{IEEEeqnarray}
Using \eqref{eqIndProd}, \eqref{denExpect} and \eqref{numExpect}, the $n$th moment of $\xi_k$ given in \eqref{XiktonExpect} can easily be derived.
\section{Integral Relationship Used in Derivation of Ergodic Capacity}
\label{sec:IntRelErgodicCap}
Consider the integral $I$ defined as
\begin{IEEEeqnarray}{c}
I = \underbrace{\int_0^{\infty} \frac{e^{-a \lambda_1 x} \ln(1+x)}{(a \lambda_1 x + b \lambda_2)^2} dx}_{I_1} +
\underbrace{\int_0^{\infty} \frac{e^{-a \lambda_1 x} \ln(1+x)}{a \lambda_1 x + b \lambda_2} dx}_{I_2}.
\end{IEEEeqnarray}
Using integration by parts, the integral $I_1$ can be expressed as
\begin{IEEEeqnarray}{c}
I_1 = \frac{1}{a\lambda_1} \int_{0}^{\infty}\frac{e^{-a\lambda_1 x}}{(1+x)(a\lambda_1 x+b\lambda_2)} dx-I_2.
\end{IEEEeqnarray}
Using partial fraction decomposition, the integral $I$ can be shown to be
\begin{IEEEeqnarray}{c}
\label{exactCapHelper}
I = \frac{e^{b\lambda_2}E_1(b\lambda_2)-e^{a\lambda_1}E_1(a\lambda_1)}{a^2 \lambda_1^2 - a b \lambda_1 \lambda_2}.
\end{IEEEeqnarray}

\bibliographystyle{IEEEtran}

\end{document}